\pdfoutput=1
\documentclass[preprint,11pt]{elsarticle}

\usepackage{graphicx}
\usepackage{subfigure}
\usepackage{hyperref}

\newcounter{bla}

\journal{Computer Physics Communications}

\newcommand{\mytitle}{PLUMED-GUI: an environment for the interactive development 
  of molecular dynamics analysis and biasing scripts}
\newcommand{\mykeywords}{Graphical User Interface \sep VMD \sep PLUMED \sep Molecular Dynamics \sep Collective Variables \sep Metadynamics}

\begin{document}

\begin{frontmatter}

\title{\mytitle}
\author{Toni Giorgino\corref{cor1}}
\ead{toni.giorgino@isib.cnr.it}
\cortext[cor1]{To whom correspondence should be addressed}
\address{Institute of Biomedical Engineering (ISIB),\\ 
National Research Council of Italy (CNR),\\
Corso Stati Uniti 4, I-35127 Padua, Italy}

\begin{abstract}
  PLUMED-GUI is an interactive environment to  develop and
  test complex PLUMED scripts within the Visual Molecular Dynamics
  (VMD) environment. Computational biophysicists can take advantage of
  both  PLUMED's  rich syntax to define
  collective variables (CVs) and VMD's chemically-aware atom selection
  language, while working within a natural point-and-click interface.
  Pre-defined templates and syntax mnemonics  facilitate the
  definition of well-known reaction coordinates. Complex CVs,
  e.g.\ involving reference snapshots used for RMSD or native contacts
  calculations, can be built through dialogs that provide a synoptic
  view of the available options.  Scripts can be either exported for
  use in simulation programs, or evaluated on the currently loaded
  molecular trajectories.   Development of scripts takes place
  without leaving VMD, thus enabling an incremental try-see-modify
  development model for molecular metrics.
\end{abstract}


\end{frontmatter}

{\bf Program summary}

\begin{small}
\noindent
{\em Manuscript Title:}                                       
 \mytitle \\
{\em Authors:}                                                
 Toni Giorgino \\
{\em Program Title:}                                          
 PLUMED-GUI (Collective variable analysis plugin) \\
{\em Journal Reference:}                                      \\
{\em Catalogue identifier:}                                   \\
{\em Licensing provisions:}                                   
 3-clause BSD Open Source. \\
{\em Programming language:}                                   
 TCL/TK. \\
{\em Operating system:}                                       
 Linux/Unix, OSX, Windows. \\
{\em RAM:}                                               
 Sufficient to run PLUMED \cite{tribello_plumed_2013} and VMD \cite{Humphrey_Dalke_Schulten_1996}. \\
{\em Number of processors used:}                              
 1 
{\em Classification:}                                         
  3 Biology and Molecular Biology, 23 Statistical Physics and Thermodynamics. \\
{\em Subprograms used:}                                       
  PLUMED (version 1.3 or higher). \\
  {\em Nature of problem:} Compute and visualize values of collective
  variables on molecular dynamics trajectories from within VMD, and
  interactively develop biasing scripts for the estimation of
  free-energy surfaces in PLUMED.
  \\
  {\em Solution method:} A graphical user interface is integrated in
  VMD and allows to interactively develop and run analysis scripts.
  Menus and dialogs provide mnemonics and documentation on the syntax
  to define complex CVs.
  \\
  {\em Restrictions:}
  Tested on systems up to 100,000 atoms. \\
  {\em Unusual features:} VMD-PLUMED is not a standalone program but a
  plugin that provides access to PLUMED's analysis features from within VMD. \\
  {\em Additional comments:} Distributed with VMD since version 1.9.0.
  Manual  update may be required  to access the latest features.   \\
  {\em Running time:} Computations of the values of collective
  variables, performed by the underlying PLUMED code, depends on the
  size of the system and the length  of the trajectory; it is 
  generally negligible with respect to simulation time.  \\

\end{small}

\section{Introduction}

Molecular dynamics (MD) is a computational technique which models the
interactions between a set of atoms with realistic empirical
potentials. Recent increases in computer power allow to routinely
sample biomolecular systems with all-atom resolution for
biologically-relevant timescales, thus providing \emph{in silico}
approximated views on processes that are too fast, or too small to be
measured \emph{in vitro}. Recent examples include protein
folding~\cite{Lindorff-Larsen_Piana_Dror_Shaw_2011}, channel
permeation and
gating~\cite{Jensen_Jogini_Borhani_Leffler_Dror_Shaw_2012}, drug
binding~\cite{Shan_Kim_Eastwood_Dror_Seeliger_Shaw_2011,Buch_Giorgino_2011},
protein-protein
interactions~\cite{Ahmad_Gu_Helms_2008,Giorgino_Buch_2012}, and so on,
not to mention applications in materials science and coarse-grained
macromolecular assemblies.

An atomistic molecular model involves thousands to millions of degrees
of freedom, which are hardly interpretable directly. Biophysically or
biochemically relevant information, such as free energies, kinetic
rates, transition probabilities, and so on,  is usually extracted
aggregating relevant degrees of freedom into reaction coordinates or
\emph{collective variables} (CVs), defined as mathematical functions of (some
of) the coordinates of the system.  CVs thus  simplify the
interpretation of complex events, and are normally used as independent
coordinates in formalisms such as the potential of mean force.

Choosing a set of CVs to adequately describe a given system is,
however, not trivial. In general, it is important to identify those
reaction coordinates which change ``slowly'' over the timescales of
the phenomena of interest. CVs thus identified can then be monitored
to detect rare events~\cite{Giorgino_Buch_2012}, be biased to
determine free energy landscapes~\cite{Laio_Parrinello_2002},  used
to partition the phase space to reconstruct
kinetic rates~\cite{Biarnes_Pietrucci_Marinelli_Laio_2012,Noe_Fischer_2008},
and so on.  Although chemical intuition is a guide in the selection of
CVs, some amount of tuning is generally required in parametrizing the
specific details of the functions.

Several software packages offer the possibility to compute CVs;
however, existing software is usually restrictive on the complexity of
the functions that can be defined, limited to the analysis phase, or
requires users to explicitly code the CV computations in ad-hoc
scripts, which therefore tend to contain ``boilerplate'' code that
obfuscates the metric. To the contrary, it would be desirable to have
a concise and human-readable definition of both the functional form
(e.g., ``distance'', ``contacts'', ``interfacial waters'', \dots) and
 the atoms involved (say, ``protein'', ``charged residues'',
``molecules close to residue X'', \dots).

A step forward in this direction is PLUMED, a flexible CV engine
recently upgraded to version 2.0~\cite{tribello_plumed_2013}. PLUMED provides
an extensive set of pre-defined \emph{actions}, i.e.\ self-explanatory
keywords that concisely define a CV on the basis of the geometry of a
system. Auxiliary actions also exist to define center of masses, ghost
atoms, units, etc.~\cite{plumed_manual,bonomi_plumed:_2009}
PLUMED scripts, in general, contain actions to define several CV,
plus, if desired, statements that express the biasing protocol to be
employed during simulation. The values of CVs can also
be computed on existing trajectories (trajectory analysis) through
its \emph{driver}  feature.

This paper introduces PLUMED-GUI, a plugin integrated with the
widely-used Visual Molecular Dynamics (VMD) molecular analysis and
visualization software~\cite{Humphrey_Dalke_Schulten_1996} to
streamline the development and test of analysis scripts.  Together,
PLUMED and PLUMED-GUI offer a concise and homogeneous way to express
CVs and evaluate them; VMD provides intuitive facilities to load and
visualize the trajectories under analysis, an easy to use graphical
environment, and a powerful, topology-aware atom selection language
for selecting molecular components.

\section{Plugin usage}

\begin{figure}
  \centering
  \subfigure[Main window]{\includegraphics[width=0.7\textwidth]{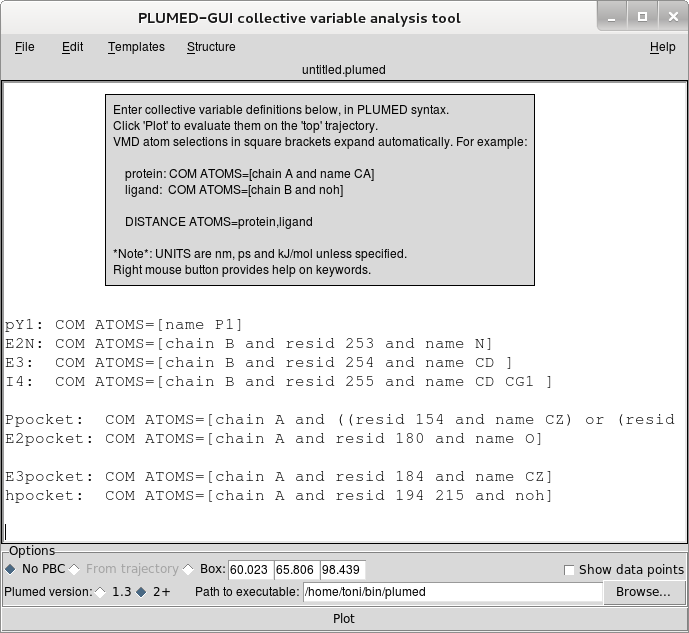} \label{fig:main}}
  \subfigure[Templates menu]{\includegraphics[width=0.23\textwidth]{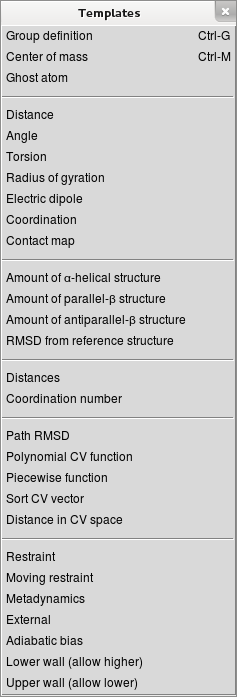} \label{fig:templates}}
  \caption{(a) PLUMED-GUI's main window.  The analysis script is
    entered in the text area, like a text editor. The \emph{Plot}
    button evaluates the collective variables defined in the script on
    the molecular trajectory currently selected in VMD (``top
    molecule''); if successful, a graph appears, showing the values of
    the CVs at each frame. The inner gray box, only shown at
    startup, is a brief reminder about the use of the interface. (b)
    The \emph{Templates} menu contains shortcuts that insert
    frequently-used definitions and collective variables. }
\end{figure}

PLUMED-GUI is started selecting the \emph{Analysis/Collective variable
  analysis (Plumed)} entry in VMD's \emph{Extensions} menu. The main
text area hosts the PLUMED script, entered following the syntax of the
PLUMED version currently in use (Figure~\ref{fig:main}).  The
interface behaves as a text editor; \emph{File} and \emph{Edit} menus
provide customary editing commands, including open and save,
copy/paste and undo/redo operations.  Initially, the text area
displays a brief syntax reminder, which can be dismissed.

It is worthwhile noting that the GUI does not restrict the input
syntax. The script is passed as-is to the underlying PLUMED engine,
with the sole exception of symbolic atom selections in square
brackets, which are resolved as will be shown in
Section~\ref{sec:symb-atom-select}.  Script coding and debugging is
entirely under the control of the user, and therefore any valid or
invalid expression can be entered.  (Consequently, the GUI needs no
updates to accommodate user-customized PLUMED variants and future
syntax.)

\subsection{Analysis and visualization}

We assume that a system of choice has been simulated by MD, and that one has
loaded the corresponding output trajectory file in VMD.
Pressing the \emph{Plot} button at the bottom of the window evaluates
the displayed script on the currently selected trajectory (known
within VMD as the all-important \emph{top} molecule). The GUI will run
PLUMED's \emph{driver} executable, which will in turn compute the values of
the CVs defined in the script at each of the top trajectory frames.

Once the evaluation is successful, the time series of the collective
variables are displayed graphically in a plot.  The purpose of the
plot is to quickly inspect the values yielded by the current CV
definitions, and provide a way to iteratively refine them. The plot
layout  shows time on the abscissa and the CV values in
different line styles; data points can be optionally read out hovering the
mouse pointer.  More complex visualizations can be obtained exporting
data to external plotting programs; data can be exported 
either as a matrix (time running as rows, and CVs as
columns), or as consecutive time-value vectors separated by empty
lines.

Should the evaluation of the script generate an error, it will be
displayed in the VMD textual console.  In most instances PLUMED
identifies the specific problem and corresponding script line; when
this happens, the error line will be highlighted as such in the text area.

\subsection{Consistency of units}

It may be worth noting that the units of computed CVs depend on
PLUMED's conventions.  Since version 2.0, PLUMED defaults to the nm,
kJ/mol, ps combination. Given that VMD users may be accustomed to the
\AA, kcal/mol, fs unit set, a reminder is shown about the fact that
the \texttt{UNITS} keyword can be used at the top of the script to
switch to customary units.

\section{Assisted script development}

\subsection{Symbolic atom selections}\label{sec:symb-atom-select}

VMD users are usually familiar with the program's powerful language
for atom selections; strings such as \texttt{same residue as (protein
  or water within 4 of name CA)} are useful expressions that are
interpreted at run time, and are equivalent to a list of atoms.  The
sophisticated syntax can query atoms on the basis of numerical
(coordinates, beta values, residue IDs), chemical (e.g.\ polar, atom
names) and/or other properties, as documented
elsewhere~\cite{Humphrey_Dalke_Schulten_1996}.

PLUMED-GUI enables the use of VMD's  selections in PLUMED scripts
through  square brackets. As shown in Figure~\ref{fig:main},
bracketed textual expressions  are evaluated with respect to
the current frame of the top molecule, and transparently replaced with the
resulting list of atoms. In this way, PLUMED users can 
avoid the use of numeric atom IDs altogether in favor of human-readable
expressions such as \texttt{[helix and name CA]}.

The use of symbolic expressions is especially advantageous when analyzing multiple
systems; this is the case, for example, when several all-atom systems
are prepared containing same protein and a series of compounds.
Whereas atom indices depend on the specific system and the details of
how it was prepared, expressions such as \texttt{[not protein
  and not water]} (matching non-peptide ligands) do not, and will be
valid regardless of the specific system being analyzed.

Symbolic atom expressions are interpreted at the moment the analysis is
started by pressing the \emph{Plot} button. They can also be
permanently replaced with atom numbers to be used independently of PLUMED-GUI, via the
\emph{Export} function (section~\ref{sec:export-use-simul}).

\subsection{Templates}

The \emph{Templates} menu provides shortcuts that insert a number of
frequently-used definitions; selecting one of the menu entries types
the corresponding keyword in the text area at the cursor's position
(Figure~\ref{fig:templates}). Templates, in other words, offer
human-readable shortcuts to enter the frequently used strings that
define atom groups and CVs. After insertion, templates can be edited
freely in the text area.
Templates have to be filled in  manually; for example, in the
case of the ``Coordination'' template, one has to specify one or two
groups between which the coordination number is to be computed, and
the parameters of the switching function.

The list of templates provided in the menu is not meant to be
exhaustive, but rather to provide a synopsis of to the most
frequently-used CVs, inserted with the default options. Generic
actions and modifiers can be typed manually, while optional keywords
can be looked up through an on-line contextual help, described
in the next section.

\subsection{On-line help}

PLUMED's actions have a wealth of options to alter the
behavior of CVs. For instance, the \texttt{COORDINATION} action
foresees modifiers to define the shape and functional form of the
switching function; to ignore periodic boundary conditions; to compute
derivatives numerically; and several others. The richness of the
syntax may make it unwieldy to recall the syntax of lesser-used
options.

\begin{figure}
  \centering
  \subfigure[Contextual help]{
    \includegraphics[width=0.45\textwidth]{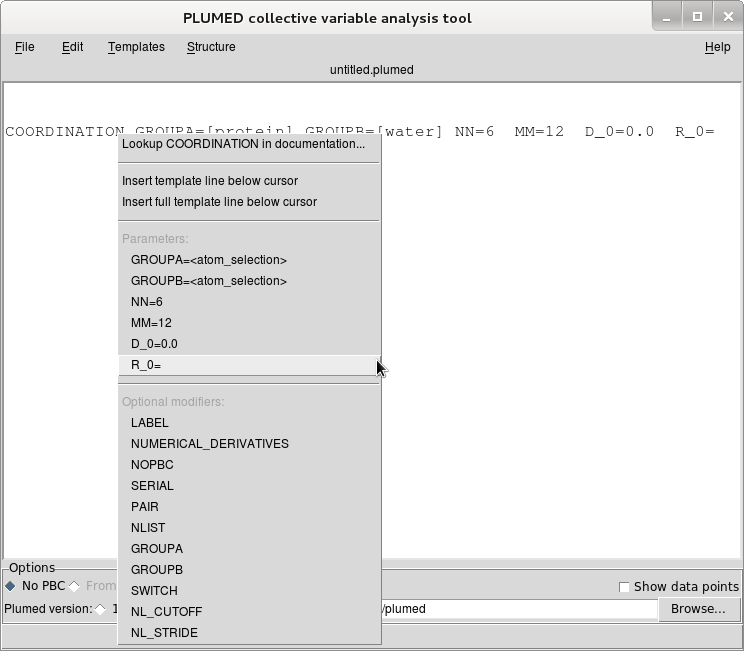}
    \label{fig:help_popup}
  }
  \subfigure[Sample manual page]{
    \includegraphics[width=0.40\textwidth]{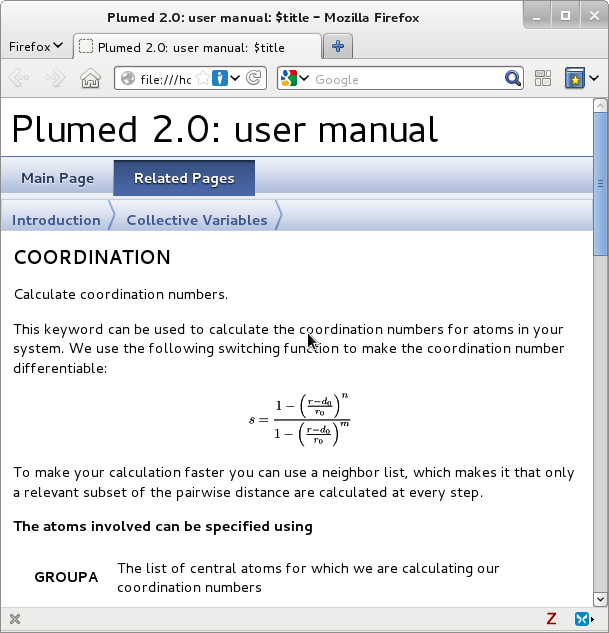}
    \label{fig:help_page}
  }    
  \caption{(a) A contextual popup menu lists mandatory and
    optional keywords supported by the action under the pointer (in
    this case, \texttt{COORDINATION}, which computes the coordination
    number of one or two groups of atoms). (b) The \emph{Lookup} function
    recalls an action's manual in the web browser. }
  \label{fig:help}
\end{figure}

To this end, PLUMED-GUI provides a comprehensive context-dependent help facility
through a pop-up menu, which is be invoked pressing the right mouse
button on any action keyword. The topmost menu item, \emph{Lookup in
  documentation}, opens up a web browser displaying the full manual
page of that action.  Subsequent entries in the pop-up menu shows the
list of optional and mandatory modifiers accepted by that
action (Figure~\ref{fig:help}).

As for the rest of PLUMED 2.0 documentation, PLUMED-GUI's contextual
help is generated automatically from PLUMED's source code.  This
implies that, as long as new features are implemented and documented
according to the established coding conventions, any newly-developed
functions become properly integrated in the interface, without
requiring modifications to the GUI code.

\section{Structure-based operations}

Functions in the \emph{Structure} menu provide assistance in the
definition of more complex CVs that depend upon the topology and coordinates of
the currently loaded system.  Each of the menu entries opens up a
dialog with a number of tunable options. Structure-based CVs
generally involve long lists of statements and/or auxiliary files;
these automated procedures are meant to relieve users from the error-prone
process of building files and lists by hand.

\begin{figure}
  \centering
  \subfigure[Build reference structure]{
    \includegraphics[scale=.45]{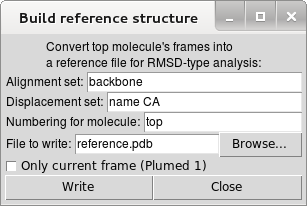} \label{fig:reference}}
  \subfigure[Native contacts]{
    \includegraphics[scale=.45]{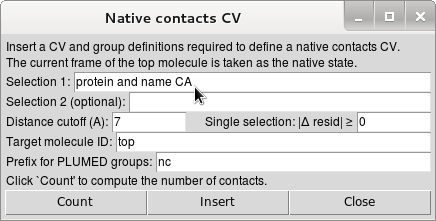} \label{fig:native}}
  \subfigure[Backbone torsion angles]{
    \includegraphics[scale=.45]{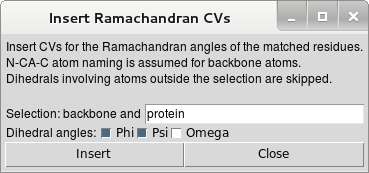} \label{fig:rama}}
  \caption{Dialogs accessible from the \emph{Structure} menu support
    the creation of CVs based on the active topology. (a) \emph{Build
      reference structure} converts the currently displayed frame into
    a reference file for RMSD calculations.  Atom sets to be used for
    alignment and displacement are specified as VMD atom selections;
    numbering can also be mapped between molecules if the reference
    frame and the trajectory on which the CV will be computed belong
    to systems with different topologies. (b) Analogously,
    \emph{Native contacts} enumerates  the atom pairs (closer
    than the chosen threshold distance) in the currently-displayed
    (``native'') frame. The CV will measure how many of
    those atom pairs will present in each trajectory frame. 
    Non-informative contacts between neighboring residues can be
    filtered out putting a lower bound to the $| \Delta \mbox{resid}
    |$ parameter. (c) The \emph{Insert backbone angles} dialog inserts
    CVs corresponding to $\phi, \psi$ and/or $\omega$ 
    dihedrals contained in the selection.}\label{fig:structure}
\end{figure}

\subsection{Generating reference structures for alignments}

The root mean square deviation (RMSD) metric is frequently used to
detect structural similarities and conformational transitions.  RMSD
values are computed averaging the squared displacement of a chosen set
of atoms (displacement set) with respect to a \emph{reference}
structure, after applying the roto-translation that optimally aligns
another, possibly coincident, set of atoms (alignment set).  PLUMED
also implements three generalization of the metric, namely the $S$,
$Z$ and \emph{property map} path variables, to express the
``progression'' and ``distance'' of the current state of the system
along a path defined by an arbitrary number of exemplary reference
structures used as
landmarks~\cite{Branduardi_Gervasio_Parrinello_2007,Spiwok_2011}.

The \emph{Build reference structure} dialog provides a convenient way
to generate such reference structures (Figure~\ref{fig:reference}).
Pressing the \emph{Write} button ``freezes'' the coordinates of the
\emph{currently selected} frame into a ``reference file''.  Reference
files are PDB-like tables used by PLUMED to define the set of atoms to
be used for alignment, for computing the displacement, and the
reference coordinates; each line represents one of the atoms involved
in the calculation, with columns recording serial numbers,
coordinates, and inclusion in one or the other
set~\cite{plumed_manual}.

The dialog allows the use of atom selections to indicate the subset of
the atoms to be involved, respectively, in the computation of the
optimal alignment, and the measure of the RMSD. A check-box provides
a choice on whether to export all of the frames of the current
trajectory (convenient when specifying a complex path), or just the
current frame (for basic RMSD calculations, or to facilitate the
manual construction of paths).

By default, the reference file generated is suitable for computing
$S$, $Z$ and property map values on systems with the same topology as the
one from which the reference was extracted.  However, it is sometimes
necessary to perform alignments between different topologies; for example, the
native structure may be a PDB file, while the system
under analysis is the all-atom structure used in
simulation. Alignments between molecules with different topologies are
possible by setting the \emph{target molecule ID}. This feature adjusts
the atom numbering of the top molecule to be compatible with the
specified \emph{target} molecule; in other words, trajectory frames of
the target molecule will be aligned with the structure of the top
molecule, even though the topologies of the two are different. The
renumbering feature requires that the atom selections match the same
number of atoms in the two systems.

\subsection{Number of native contacts}

The number of native contacts is another metric  to
determine structural similarity, frequently used as an indicator of folding
or binding.  The metric puts the accent on the presence of those
contacts that characterize the desired (native) structure. First, the
pairs of atoms in contact in a given native structure are
enumerated. Then, this list is evaluated
for each of the trajectory frames under analysis: the 
CV counts how many of the pairs that were in contact in the
reference frame are also close in the frame being analyzed.  

The \emph{Native contacts} dialog (Figure~\ref{fig:native}) can be
used to generate such lists flexibly and with ease. Like when building
reference structures, the current frame of the top molecule is used as
the native state.  It is possible to specify either one or two atom
selections; in the first case, the contacting pairs involving atoms in
the selection are enumerated; otherwise, if two selections are given,
intermolecular contacts -- bridging the two selections -- will be
counted.  The ``distance cutoff'' box adjusts the distance (in \AA) at
which an atom pair is assumed to be in contact.

A marked rise in the number of native contacts is often used as a
proxy for the detection of folding events. However, residues adjacent
in the primary sequence will almost always be in contact, thus
contributing little or no information to the folding signal. These
``trivial'' contacts can be filtered out setting a minimum bound to
the $| \Delta \mbox{resid} |$ to a positive integer $d$. If set,
contacts between atoms closer than $d$ residues apart in the
primary sequence will be disregarded.
Analogously to the \emph{Build reference structure} function, the user
can match a trajectory with a native frame with a different topology
specifying the appropriate target molecule ID.

The number of native contacts is implemented in PLUMED through the
\texttt{COORDINATION PAIRS} action and the enumeration of the
contacting pairs in the native frame.  It is worthwhile noting that,
like all other CVs provided by PLUMED, this metric is a continuous
approximation of the integer pair count, made smooth with respect to
all of the system's coordinates through an exponential switching
function~\cite{plumed_manual}.

\subsection{Backbone torsion angles}

The \emph{Insert backbone angles} dialog (Figure~\ref{fig:rama})
allows the computation of backbone $\phi$, $\psi$ and/or $\omega$
torsion angles between neighboring residues, defined according to the
standard IUPAC rules for biochemical nomenclature~\cite{IUPAC}.  The
user is asked to specify an atom selection; when the \emph{Insert}
button is pressed, a CV will be inserted for each $\phi$, $\psi$
and/or $\omega$ backbone dihedral contained in the selection. Each
angle is defined through the appropriate \texttt{TORSION} keyword and,
for the sake of readability, includes a comment pointing back to the
name of the involved residue.

\section{Export for use in simulation}\label{sec:export-use-simul}

PLUMED has extensive facilities to biases molecular dynamics
simulations with forces that enhance the sampling of the phase-space
in a way that allows the reconstruction of free-energy
surfaces. Example of biasing protocols include harmonically
constraining CVs at a given combination of values (used e.g.\ for the
umbrella sampling protocol \cite{Torrie_Valleau_1977}), pulling them
towards increasing or decreasing values (steered MD
\cite{Isralewitz_Gao_Schulten_2001,Giorgino_2011}), metadynamics
\cite{Laio_Parrinello_2002}, and so on. Biased MD simulations are
carried out with codes patched to embed the PLUMED engine. Force
biases are specified in the script, which defines the  protocol
as well as the CVs to be biased.\footnote{A tutorial on biasing and
  the search of CVs suitable for
  specific biomolecular systems can be found  e.g.\ at
  \url{www.plumed-code.org/documentation}.}  Atoms have to be
specified through their serial numbers, which makes the iteration of
complex scripts through different systems an error-prone exercise.

The \emph{Export} function, accessible from the \emph{File} menu, removes all
the symbolic atom selections in the current script and replaces them
with the corresponding numerical lists. The exported script is thus
devoid of VMD-specific constructs, and can then be employed for
simulations.  The exported file contains comments to document how the
numeric atom lists were obtained although, for the sake of
reproducibility, it is generally advisable to keep the original
script with unsubstituted, symbolic atom selections.

\section{Installation and compatibility}

The GUI supports the same wide range of platforms as VMD, encompassing
all major variants of Linux/Unix, OSX, and Windows.  Trajectory
analysis is performed invoking the platform-specific \emph{driver}
executable behind the scenes. PLUMED distributions provide
instructions on how to build the executable on Unix-like systems; 
a precompiled version for Windows is available for convenience, which
can downloaded and installed automatically.

The current version of the plugin, PLUMED-GUI 2.2, supports both
PLUMED 1.3 and PLUMED 2.0, with minor functional differences.  For
clarity, this manuscript focused on the features available when using
PLUMED version 2.0 as a back-end.  Language syntax and \emph{driver}
invocation method differ between the two PLUMED versions.  The GUI
detects which version is installed and adapts templates and syntax
accordingly.  If both PLUMED versions are available, the user can
switch manually between the two.

Recent VMD distributions contain a preinstalled version of PLUMED-GUI.
Users may download the latest version and supporting material from the
address \url{www.multiscalelab.org/utilities/PlumedGUI}.

\section{Conclusions}

Developing an appropriate combination of reaction coordinates is a
central task in the analysis of biomolecular systems.  PLUMED-GUI
simplifies the iterative development, refinement and test of
collective variables to be used with the PLUMED engine.  The GUI
bridges the usability of VMD's graphical interface and PLUMED's rich
CV definition language.  

Integrating the two environments incurs in a few limitations; right
now, only orthorhombic simulation boxes with constant edges are
supported, therefore precluding the analysis of constant-pressure
simulations (this limitation may be removed as soon as \emph{driver}'s
support to trajectory formats is expanded). Another drawback is due to
the fact that atom selections are evaluated only once, before the
computation is started; thus, it is not possible to employ
time-varying atom lists (nor PLUMED engine would support them):
analysis protocols involving time-varying atom sets are outside of the
scope of the programs. It is worthwhile noting, however, that PLUMED~2
provides switching functions (such as \texttt{DISTANCES LESS\_THAN})
that are continuous approximations to discrete quantities
such as the number of atoms satisfying a given property.

One of the objectives of PLUMED-GUI is to lower the barrier for the
adoption of meaningful metrics in the analysis tasks of simulation
data. In the future, the interface may be expanded integrating more
``function building'' features and providing interfaces with external
programs, such as METAGUI~\cite{Biarnes_Pietrucci_Marinelli_Laio_2012}
and reweighting schemes~\cite{Bonomi_Barducci_Parrinello_2009}.

\section{Acknowledgments}

I would like to thank the authors of PLUMED and VMD for creating,
distributing and supporting the corresponding software packages. An
acknowledgment goes to Prof.\ G.\ De Fabritiis and his group at the
Computational Biophysics Laboratory at the Universitat Pompeu Fabra
(Barcelona), where this work was started. Former support from the
Ag\`encia de Gesti\'o d'Ajuts Universitaris i de Recerca, Generalitat
de Catalunya (2009 BP-B 00109) is gratefully acknowledged.

\bibliographystyle{model1-num-names}

\vspace{2cm}

\end{document}